\documentclass[%
 aip,
 amsmath,amssymb,
 reprint,%
floatfix
]{revtex4-1}

\usepackage{graphicx}
\usepackage{dcolumn}
\usepackage{bm}

\usepackage[utf8]{inputenc}
\usepackage[T1]{fontenc}
\usepackage{mathptmx}
\usepackage{etoolbox}
\usepackage{comment}
\usepackage{amsmath}
\DeclareMathOperator{\sgn}{sgn}
\DeclareMathOperator{\erf}{erf}
\makeatletter
\def\@email#1#2{%
 \endgroup
 \patchcmd{\titleblock@produce}
  {\frontmatter@RRAPformat}
  {\frontmatter@RRAPformat{\produce@RRAP{*#1\href{mailto:#2}{#2}}}\frontmatter@RRAPformat}
  {}{}
}%
\makeatother

\begin{document}

\preprint{AIP/123-QED}

\title{Full counting statistics of ultrafast quantum transport}
\author{M. Hübler}
\author{W. Belzig}%
\email{Wolfgang.Belzig@uni-konstanz.de, Matthias.Huebler@uni-konstanz.de}
\affiliation{ 
Fachbereich Physik, Universit\"at Konstanz, D-78457 Konstanz, Germany
}%

\date{\today}

\begin{abstract}
Quantum transport in the presence of time-dependent drives is dominated by quantum interference and many-body effects at low temperatures. For a periodic driving, the analysis of the full counting statistics revealed the elementary events that determine the statistical properties of the charge transport. As a result, the noise correlations display quantum oscillation as functions of the ratio of the voltage amplitude and the drive frequency  reflecting the detailed shape of the drive. However, so far only continuous wave excitations were considered, but recently transport by few-cycle light pulses were investigated and the need for a statistical interpretation became eminent. We address the charge transfer generated by single- or few-cycle light pulses. The fingerprints of these time-dependent voltage pulses are imprinted in the full counting statistics of a coherent mesoscopic conductor at zero temperature. In addition, we identify the elementary processes that occur in the form of electron-hole pair creations, which can be investigated by the excess noise. We study the quantum oscillations in the differential noise induced by a wave packet consisting of an oscillating carrier modulated by a Gaussian- or a box-shaped envelope. As expected, the differential noise exhibits an oscillatory behavior with increasing amplitude. In particular, we find clear signature of the so-called carrier-envelope phase in the peak heights and positions of these quantum oscillations. More carrier cycles under the Gaussian envelope diminish the influence of the carrier-envelope phase, while this is not true for the box pulses, probably related to the abrupt onset. As the quantum oscillations are due to the energy-time uncertainty of the short pulses, our results pave the way to a description of the nonequilibrium electron transport in terms of a many-body Wigner function of the electronic system.
\end{abstract}

\maketitle

\section{Introduction}
Oscillating electric fields impact the electron transport in mesoscopic conductors, which leads to features at multiples of the driving frequency \cite{tien:1963,LESOVIK:1994tq,Schoekopf:1998}. A mesoscopic conductor divides the incident electron stream from the terminals according to its total transmission probabilities \cite{Blanter:2000wi}. These charge carriers are distributed by the equilibrium Fermi function, i.e., by the chemical potentials and temperatures of their departure terminals. Photon-assisted transport describes electron tunneling driven by an oscillating chemical potential \cite{Buettiker:1994,Pedersen:1998uc}. The electron wave function at energy $E$ exhibits side bands at energies $E+n \hbar \omega, n \in \mathbb{Z}$ weighted by Bessel functions \cite{Buettiker:1994,Pedersen:1998uc,Blanter:2000wi}. This leads to kinks at voltages $n \hbar \omega/e$ in the zero-temperature noise. The step height in the differential noise depends on the ac amplitude through the Bessel functions \cite{LESOVIK:1994tq,Blanter:2000wi}. This field is complemented by the dynamical response of the shot noise \cite{Gabelli:2008,Hammer:2011bl}.

Going beyond the noise and analyzing the full counting statistics (FCS) gives insight into the elementary processes involved in electron transport \cite{levitov:93,Levitov:1996,Belzig:2001kw,DiLorenzo:2005kz}. Classical charge carriers would obey a Poisson distribution, while electrons in a mesoscopic conductor follow a binomial distribution due to the Pauli exclusion principle \cite{levitov:93,Levitov:1996,Ivanov_Lee_Levitov_1996}. This manifests itself in a suppression of noise, indicated by the Fano factor. In superconducting junctions, the probability distribution can assume negative values, which is inconsistent with the notion of a probability distribution \cite{Belzig:2001kw}. Nevertheless, this function accounts for all elementary processes and can in principle be measured in an experiment. The information about the charge transport is contained in the cumulant generating function (CGF). Dismantling the CGF into a sum of binomials or multinomials unveils the independent elementary processes at play. This shows, for instance, that the charge transfer in spin-sensitive tunnel junctions is carried by single electrons and spin-singlet pairs when the electron source is a mesoscopic conductor in series \cite{DiLorenzo:2005kz}.

Time-dependent voltage pulses generate collective excitations of the Fermi sea, which are realized as a superposition of electron and hole excitations \cite{Levitov:1996,Ivanov:1997wz,Vanevic:2016eq}. For example, Levitons are soliton-like electron excitations, created by Lorenzian voltage pulses with a quantized flux \cite{Levitov:1996,Ivanov:1997wz}. These states minimize the noise, which is reduced to a corresponding dc noise. Furthermore, the entire full counting statistic shortens to a corresponding dc statistic \cite{Vanevic:2007id,Vanevic:2008tf}. Levitons have been realized experimentally \cite{Dubois:2013} and analyzed via quantum state tomography \cite{jullien:2014}. In general, many body electronic states generated by a time-dependent voltage are built from single-electron and electron-hole quasiparticle excitations \cite{Gabelli_Reulet_2013,Vanevic:2016eq}. The amplitudes and probabilities of these elementary excitations depend on the details of the drive and can be examined experimentally \cite{Vanevic:2016eq}. The entanglement can be addressed by a continuous-variable entanglement test \cite{Zhan:2019bo}.

Arbitrary time-dependent voltages can be treated by the non-equilibrium Green's function method \cite{nazarov:1999,Belzig:2001kw,Vanevic:2007id,Vanevic:2008tf}. The work \cite{Vanevic:2007id,Vanevic:2008tf} investigated the FCS for cosine, square, triangle, sawtooth, and Lorentzian pulses. Apart from a dc statistic, there appears to be a bidirectional part that depends on the scattering of electron-hole pairs. The probabilities of creating these electron-hole pairs enter the CGF and are determined by the shape of the pulses. A similar CGF structure is found for a superconductor (S)-normal-metal (N) contact, because the problem can be mapped onto a NN contact by replacing the transmission probabilities with the Andreev reflection probabilities and doubling the counting field as well as the applied voltage \cite{Belzig:2016jz}.

Nowadays, light fields can be manipulated on femtosecond time scales, leading to ultrafast electron transport in mesoscopic constrictions. The experiments \cite{Rybka:2016hx,ludwig:2019,Ludwig:2020} harnessed the nonlinear $I-V$ characteristic of their bow-tie antenna to establish carrier-envelope phase (CEP) control of electron transfer. For a cosine-shaped carrier to envelope configuration (CEP$=0$), the average current is maximal, while for a sine-shaped configuration (CEP$=\pi/2$), the average current vanishes. The near-field in the gap is distorted compared to the far-field, leading to a strong field enhancement and a shift in the effective carrier-envelope phase \cite{ludwig:2019}. A small dc-bias can be exploited to direct the photocurrent \cite{Ludwig:2020}.

Our work extends the investigations \cite{Vanevic:2007id,Vanevic:2008tf} to include pulses that are only a few cycles long and modulated by a Gauss- or box-shaped envelope. We are interested in the zero-temperature noise and how the carrier-envelope phase manifests therein. As expected, the differential noise shows oscillations with increasing ac amplitude. For both envelopes, the carrier-envelope phase clearly affects these oscillations. The influence of the carrier-envelope phase attenuates for rising number of carrier cycles under the Gauss. This trend is not seen for the box pulse, probably because the temporal behavior at the edges of the box is fundamentally different for a sine and cosine carrier. 

The article is structured as follows: In the second section \ref{sec:methods}, we recapitulate the methods \cite{Vanevic:2007id,Vanevic:2008tf} dealing with the FCS in the context of time-dependent voltages. The third second \ref{sec:results} is concerned with the results for the Gauss and box pulses. The last Section \ref{sec:conclusion} summarizes our findings.

\section{Methods}
\label{sec:methods}
The full counting statistic is concerned with the probability of charge transfer \cite{levitov:93,Levitov:1996}. The transfer of $N$ charges in a given measurement time $t_0$ is described by the probability density $P_{t_0}(N)$. For uncorrelated and unidirectional transport, we get a Poissonian distribution that exhibits Schottky's shot noise. In contrast, for a degenerated electron gas at zero temperature, we obtain a binomial distribution.

All information about the cumulants is encoded in the cumulant generating function (CGF) $\mathcal{S}(\chi)$, which is related to the probability density by
\begin{align}
    \exp(-\mathcal{S}(\chi))=\sum_N P_{t_0}(N) \exp(i N \chi),
\end{align}
with the counting field $\chi$, and the imaginary unit $i$ \cite{levitov:93,Levitov:1996}. The normalization of the distribution requires $\mathcal{S}(0)=0$. The cumulants are attained by
\begin{align}
    C_n=-(-i)^n \frac{\partial^n}{\partial \chi^n} \mathcal{S}(\chi) \bigg |_{\chi=0},
\end{align}
where the first cumulant $C_1=\overline{N} := \sum_N N P_{t_0}(N)$ is the mean value, the second cumulant $C_2=\overline{(N-\overline{N})^2}$ the noise. Expressed by the current operator $\hat{I}(t)$ at time $t$, the first cumulant corresponds to
\begin{align}
    C_1=\frac{1}{-e} \int_0^{t_0} dt \langle \hat{I}(t) \rangle
\end{align}
and the second cumulant to
\begin{align}
    C_2=\frac{1}{2 e^2} \int_0^{t_0} \int_0^{t_0} dt dt' \langle \{\Delta \hat{I}(t),\Delta \hat{I}(t') \} \rangle,
\end{align}
with $\Delta \hat{I}(t)=\hat{I}(t)-\langle \hat{I}(t) \rangle$, $\{\cdot,\cdot\}$ the anticommutator and $e$ the elementary charge. 

In a coherent mesoscopic conductor, the CGF depends on an ensemble of transmission eigenvalues $\{T_n \}$ and on the quasiclassical Green's functions of the terminals \cite{nazarov:1999,Belzig:2001kw}. The Green's function is determined by the Fermi distribution, i.e., by the chemical potential and the temperature of the corresponding terminal. If we apply a dc bias between the terminals, then each energy contributes separately to the CGF, where the bias enters the Fermi distribution as the difference between the chemical potentials. 

Time-dependent voltages $V(t)$ correlate transport at different energies, and the CGF depends on their entire shape \cite{Vanevic:2007id,Vanevic:2008tf,Belzig:2016jz}. We include time-dependent voltages via a gauge-like transformation of one Green's function, similar to the counting field. The transformed Fermi function has no longer a diagonal structure in energy representation. The CGF does not decouple between different energies and remains a complex convolution sum. In the zero temperature limit, this can be solved for periodic voltages by a diagonalization procedure, described in detail in \cite{Vanevic:2007id,Vanevic:2008tf}. The CGF $\mathcal{S}(\chi)=N_{\text{uni}} \mathcal{S}_{\text{uni}}(\chi)+N_{\text{bi}1} \mathcal{S}_{\text{bi}1}(\chi)+N_{\text{bi}2} \mathcal{S}_{\text{bi}2}(\chi)$ subdivides into a unidirectional part 
\begin{align}
    \mathcal{S}_{\text{uni}}(\chi)=\sum_n \ln[1+T_n(e^{-i \kappa \chi}-1)].
\label{eq:unidirectional}
\end{align}
and bidirectional parts
\begin{align}
    \mathcal{S}_{\text{bi}1,2}(\chi)= \sum_n \sum_k \ln[1- T_n R_n p_{k1,2} (e^{i \chi}-1)(e^{-i \chi}-1) ].
\label{eq:bidirectional}
\end{align}
The numbers of attempts in the measurement time are given by $N_{\text{uni}}=t_0|e \overline{V}|/\pi $, $N_{\text{bi}1}=t_0 \Omega_1/\pi$, and $N_{\text{bi}2}=t_0(\Omega-\Omega_1)/\pi$, where $\Omega_1=e\overline{V}-\lfloor e \overline{V}/\Omega \rfloor \Omega$, with $\lfloor x \rfloor$ the floor of the real number $x$. In the formulas, the repetition period $\tau=2\pi/\Omega$ of the pulses, the average voltage $\overline{V}=(1/\tau) \int_{-\tau/2}^{\tau/2} V(t) dt$, the probabilities $p_{k1,2}$ of creating an electron-hole pair, and the reflection probabilities $R_n=1-T_n$ occur. The parameter $\kappa=1$ ($\kappa=-1$) indicates the direction of charge transfer, determined by $e \overline{V}>0$ ($e\overline{V}<0$). We set $\hbar=1$ in all expressions.

Unidirectional transport \eqref{eq:unidirectional} corresponds to a single electron transfer with probability $T_n$, where each transport channel contributes independently. Within the measurement time $t_0$, there are $t_0 |e\overline{V}|/\pi$ attempts of electrons to cross the conductor. Hence, unidirectional charge transfer only occurs at finite average voltages $\overline{V}$. The period $\tau$ and the characteristic correlation time of the current fluctuations are assumed to be much smaller than the measurement time. For a single transport channel, the CGF translates to a binomial distribution. 

Bidirectional charge transport \eqref{eq:bidirectional} stems from two particle processes, namely the creation of electron-hole pairs in one terminal by the ac voltage $\Delta V(t)=V(t)-\overline{V}$. This pair contributes to the statistics if the electron traverses the conductor and the hole gets reflected, or vice versa. This process is captured by the probability $T_n R_n p_k$, which is composed of the transmission probability $T_n$, the reflection probability $R_n$, and the probability $p_k$ of creating an electron-hole pair. The index $k$ labels different electron-hole pair excitations of the Fermi sea \cite{Vanevic:2016eq}. We have to encounter two types of bidirectional processes. They differ by the number of attempts $N_{\text{bi}1,2}$ and electron-hole pair creation probability $p_{k1,2}$. In contrast to the unidirectional charge transport, the bidirectional parts only contribute to the noise and higher-order even cumulants.

The details of the time-dependent drive determine the probabilities of electron-hole pair creation. We obtain the probabilities $p_k$ by diagonalizing the matrix
\begin{align}
    M_{nm}(E)=\sgn(E+n \Omega) \sum_{k=-\infty}^{\infty} a_{n+k} a^*_{m+k} \sgn(E-k \Omega-e \overline{V}),
    \label{eq:mat}
\end{align}
with the signum function $\sgn(\cdot)$, $a^*_n$ the complex conjugate of $a_n$ and the coefficients
\begin{align}
    a_n=\frac{(-1)^n}{\tau} \int_{-\tau/2}^{\tau/2} dt \exp \left(-i \int_{-\tau/2}^t dt' e \Delta V(t') \right) e^{in \Omega t}.
    \label{eq:coeff}
\end{align}
The coefficients depend on the ac drive $\Delta V(t)=V(t)-\overline{V}$ and obey the properties
\begin{align}
    \sum_{k=-\infty}^\infty a_{n+k} a^*_{m+k}=\delta_{nm}, \, \, \, \, \sum_{k=-\infty}^\infty k |a_k|^2=0.
\end{align}
In the relevant energy range $0<E<\Omega$, the matrix is piecewise constant for energies $E \in (0,\Omega_1)$, $\Omega_1=e\overline{V}-\lfloor e \overline{V}/\Omega \rfloor \Omega$ and $E \in (\Omega_1,\Omega)$. This leads to the two types of bidirectional processes denoted by an index of $1$ for the first interval and an index of $2$ for the second interval. The matrix $M_{nm}$ is unitary, and therefore the eigenvalues assume the form $\exp(\pm i \alpha_{k1,2})$. If the unidirectional number $e \overline{V}/\Omega$ of attempts per cycle and per spin is an integer, then there is only one bidirectional CGF $\mathcal{S}_{\text{bi2}}(\chi)$. The phase $\alpha_{k1,2}$ of the eigenvalues enters directly into the electron-hole pair creation probability by $p_{k1,2}=\sin^2(\alpha_{k1,2}/2)$. The corresponding eigenvectors are related to the electron-hole quasiparticle amplitude and dictate the drive induced many-body wave function \cite{Vanevic:2016eq}.

The bidirectional current-current noise is proportional to the sum of all probabilities $p_k$, which successively rise to one as the amplitude of the ac-drive is increased. In the following, we are concerned with time dependent voltages, which exhibit a vanishing average voltage $\overline{V}=0$. Hence, the unidirectional noise vanishes and only type $2$ bidirectional processes appear. The current-current noise per pulse and per spin has the form
\begin{equation}
    S=2e^2 \left(\sum_n T_n R_n \right) \sum_k p_k.
\end{equation}
Based on the findings in \cite{Vanevic:2007id,Vanevic:2008tf}, we expect oscillatory behavior in the differential noise $\partial S/(\partial (eV_0))$, which fades out for large ac amplitudes. The oscillations come from the successive ramp up of the probabilities $p_k$ with ac amplitude, where the slope determines the height and the turning point the position of the peak.
\section{Results}
\label{sec:results}
The considered voltage pulses consist of an oscillating carrier modulated by an envelope. Their displacement against each other is described by the carrier-envelope phase (CEP). We are after the effect of the CEP on the noise. The voltage assumes the form
\begin{equation}
    V(t)=E(t) V_0 \cos(\omega t+\phi)-V_{\text{off}},
\end{equation}
with the envelope $E(t)$, the ac amplitude $V_0$, the carrier frequency $\omega$, the carrier-envelope phase $\phi$, the offset $V_{\text{off}}=(1/\tau)\int_{-\tau/2}^{\tau/2} E(t) V_0 \cos(\omega t+\phi)$, and the repetition rate $1/\tau$. The offset is subtracted to ensure a vanishing average voltage.

For the following results, we calculated the coefficients \eqref{eq:coeff} and diagonalized the matrix \eqref{eq:mat} to obtain the electron-hole pair creation probabilities. The sum of the probabilities gives the noise. Numerically, we utilize a finite-dimensional matrix to obtain the eigenvalues. Cutoffs for $n$ and $m$ are chosen on a much larger scale than the one on which $|a_n|$ vanishes. 

\begin{figure}[tb]
    \centering
    \includegraphics{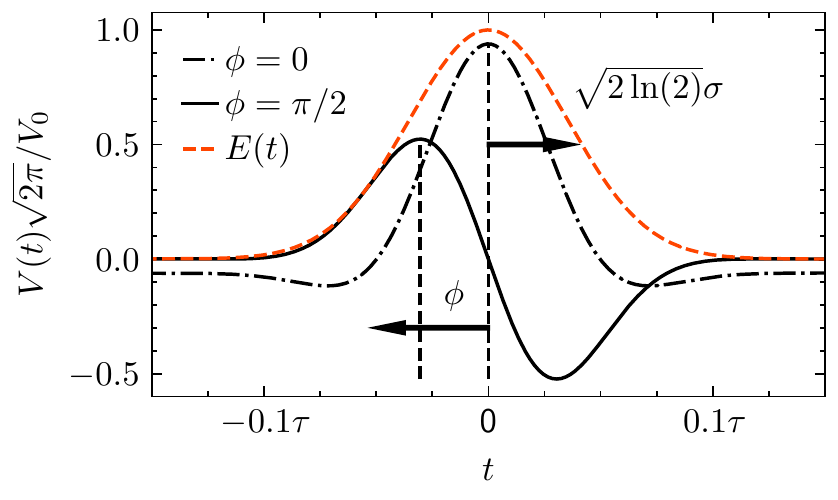}
    \caption{Gauss pulse and its envelope for carrier-envelope phases $\phi=0,\pi/2$. The ratio between the variance $\sigma$ and the pulse repetition time $\tau$ is chosen as $\tau/(\sqrt{2} \sigma)=10$. Here, we set $\sigma \omega=1$, which is a measure of the number of carrier cycles under the Gauss curve. Note the negative offset voltage for $\phi=0$ to keep the average voltage at zero.}
    \label{fig:GaussZeroPuls}
\end{figure}

The first envelope we choose is a Gauss curve of the contour 
\begin{equation}
    E(t)=\frac{1}{\sqrt{2\pi}}\exp \left(-\frac{t^2}{2 \sigma^2} \right),
\end{equation}
where $\sigma^2$ corresponds to the variance. An illustration of the Gaussian pulse is depicted in Fig. \ref{fig:GaussZeroPuls}. 
The envelope shape is controlled by the ratio $\tau/(\sqrt{2} \sigma)$, which determines the relation between the width of the Gauss and the repetition rate. A measure for the number of carrier cycles under the envelope constitutes $\sigma \omega$. The offset voltage is given by
\begin{align}
    V_{\text{off}}=&\cos \phi \frac{\sigma}{\tau} V_0 e^{-(\sigma \omega)^2/2}  \Re  \bigg [ \erf \left(\frac{\tau}{2 \sqrt{2} \sigma}-\frac{i \sigma \omega}{\sqrt{2}} \right) \bigg], 
\end{align}
with the error function $\erf (\cdot)$, and the real part $\Re [\cdot]$ of a complex number. A maximum is achieved for $\phi=0$ and a minimum for $\phi=\pi/2$.\\
\\
The CEP has an observable impact on the differential noise $\partial S/(\partial (eV_0))$, depicted in Fig. \ref{fig:GaussZeroPuls} for carrier-envelope phases between $\phi=0$ and $\phi=\pi/2$.
\begin{figure}[tb]
    \centering
    \includegraphics{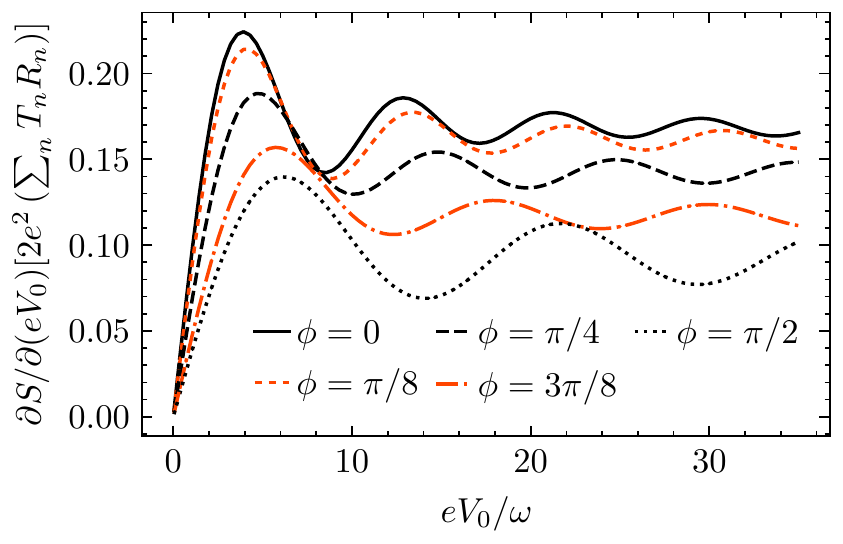}
    \caption{The differential noise $\partial S/(\partial (eV_0))$ of a Gaussian pulse is depicted over the ac amplitude $V_0$ for carrier-envelope phases between $0$ and $\pi/2$. The variance $\sigma$ in relation to the pulse repetition time $\tau$ was set to $\tau/(\sqrt{2}\sigma)=10$ and in relation to the carrier angular frequency $\omega$ to $\sigma \omega=1$.}
    \label{fig:Diff_GaussZero_phi}
\end{figure}
Successive ramp ups of the electron-hole pair creation probabilities $p_k$ cause the oscillating character of the differential noise. The CEP has a noticeable influence on the onset and slope of these ramp ups. This is reflected in the differential noise by the position and height of the peaks. For example, $\phi=\pi/2$ exhibits a larger period and smaller height compared to $\phi=0$.

\begin{figure}[tb]
    \centering
    \includegraphics{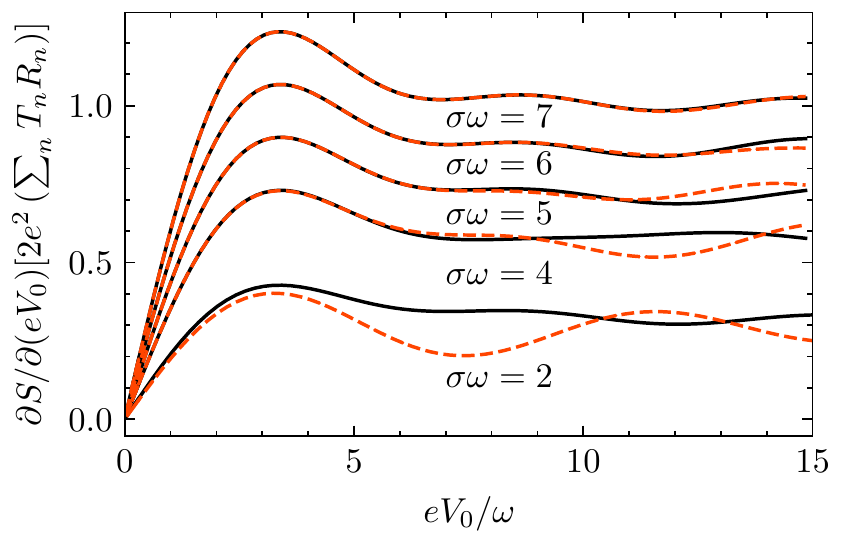}
    \caption{Differential noise $\partial S/(\partial (eV_0))$ of a Gaussian pulse for different numbers of carrier cycles under the Gaussian envelope. We compare the carrier-envelope phase $0$ (black solid) to $\pi/2$ (orange dashed) for $\sigma \omega=2$ to $\sigma \omega=7$. As the number of carrier cycles increases, the curves with CEP $0$ and $\pi/2$ fall on top of each other. Additionally, more elementary processes $p_k$ are contributing, and this leads to an increase in the differential noise. The ratio $\tau/(\sqrt{2}\sigma)$ was fixed at $10$.}
    \label{fig:Diff_GaussZero_omega}
\end{figure}

The influence of the CEP diminishes with the increasing number of carrier cycles in a pulse. Figure \ref{fig:Diff_GaussZero_omega} outlines the dependence of the differential noise on the number of carrier cycles and their influence on the CEP dependence. 
 For $\sigma \omega=2$ the curves for $\phi=0$ and $\phi=\pi/2$ are significantly different, while for rising $\sigma \omega$ they approach each other, and for $\sigma \omega=7$ they almost overlap. An additional effect is that more probabilities ramp up at the same amplitude $eV_0/ \omega$, which shifts the oscillations upward. In comparison to $\sigma \omega=1$ (see Fig. \ref{fig:Diff_GaussZero_phi}), the second peaks are shifted to lower amplitudes.

The pulses ought to be well separated, i.e., $\tau/(\sqrt{2}\sigma)\gg 1$. In this regime, the differential noise is ideally independent of this parameter. We chose $\tau/(\sqrt{2}\sigma)=10$ as a trade-off between the computational costs and the already independent probabilities in the case of $\phi=\pi/2$. For $\phi=0$, the probabilities slightly change with increasing $\tau/(\sqrt{2}\sigma)>10$, presumably due to the non-negligible influence of the offset voltage. However, the differences in the oscillations are not exclusively attributable to the offset voltage, because we observe an even smaller period and higher peaks for values $\tau/(\sqrt{2}\sigma)>10$.

\begin{figure}[tb]
    \centering
    \includegraphics{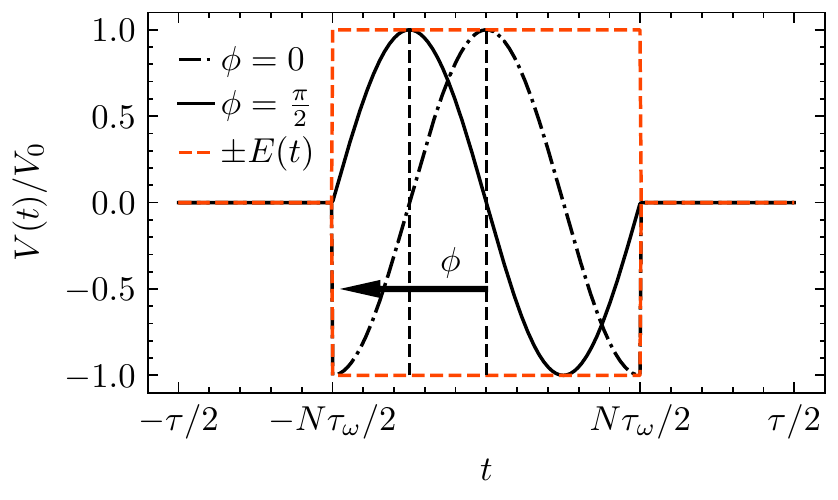}
    \caption{Illustration of the box pulse and its envelope for carrier-envelope phases $\phi=0,\pi/2$. The box counts $N=1$ carrier cycles with a period of $\tau_\omega$ and extends between $-N \tau_\omega/2$ and $N \tau_\omega/2$, which is assumed to be smaller than the pulse repetition time $\tau$.}
    \label{fig:BoxPuls}
\end{figure}

As the second envelope, we study a box of the form
\begin{align}
    E(t)=\Theta(t+N \tau_\omega/2)(1-\Theta(t-N\tau_\omega/2)),
\end{align}
with $\Theta(\cdot)$ the Heaviside function, $N$ the number of carrier cycles in the box and $\tau_\omega=2 \pi/\omega$ the period of one carrier cycle. The offset voltage vanishes for all carrier-envelope phases.
Figure \ref{fig:BoxPuls} depicts the box pulse for CEP $\phi=0,\pi/2$.

\begin{figure}[tb]
    \centering
    \includegraphics{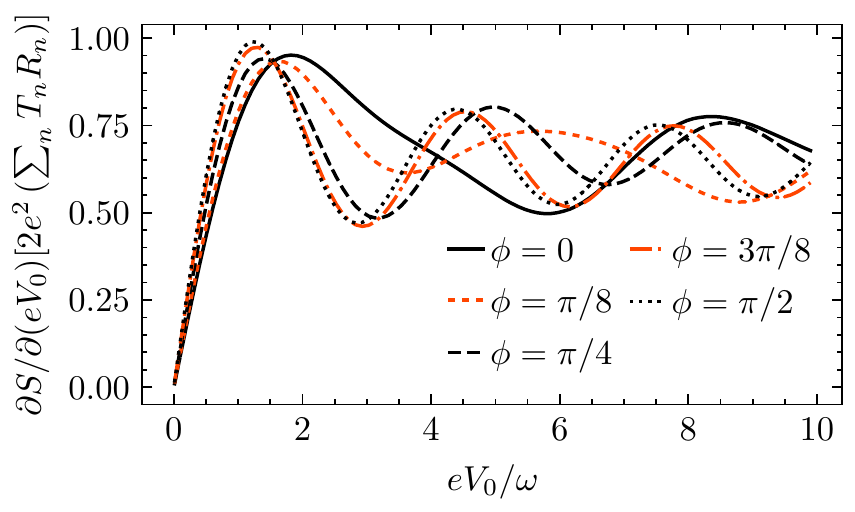}
    \caption{Differential noise $\partial S/(\partial (eV_0))$ of a box pulse drawn over the ac amplitude $V_0$ for carrier-envelope phases between $0$ and $\pi/2$. The number of carrier cycles is $N=1$. The width of the box in comparison to the pulse repetition rate was fixed to $\tau/(N\tau_\omega)=2$, because the electron-hole pair creation probabilities are nearly independent of the box size. Oscillations are caused by elementary processes that are subsequently activated as the voltage amplitude increases.}
    \label{fig:Diff_Box_phi}
\end{figure}

Again, traces of the CEP are pronounced in the differential noise  $\partial S/(\partial (eV_0))$. The dependence on different carrier-envelope phases is shown in Fig. \ref{fig:Diff_Box_phi}.
It impacts the position and slightly the height of the peaks. For example, for $\phi=0$, the second peak almost forms an excess wing of the first peak, while for $\phi=\pi/2$ the second peak is clearly visible. 

With increasing carrier cycles $N$ in the box, more probabilities $p_k$ ramp up at the same ac amplitude. Therefore, the differential noise experiences an upward shift, akin to the Gaussian envelope. In contrast to the Gaussian pulse, the peak positions remain unchanged for the first few $N$ investigated. The differential noise for $\phi=0$ and $\phi=\pi/2$ do not approach each other and stay distinct. We assume that the origin of this lies in the different behavior of the cosine and sine at the edges of the box. The cosine ($\phi=0$) jumps to zero, and the sine ($\phi=\pi/2$) is zero but exhibits a kink. 

\section{Discussion}
\label{sec:conclusion}
We have addressed the statistical properties of the current driven by few-cycle voltage pulses at low temperature in a mesoscopic conductor. Explicitly, we have studied single- and few-cycle pulses with a Gaussian envelope and with a box envelope. The noise is carried by bidirectional processes that lead to an oscillatory behavior of the differential noise. These oscillations strongly depend on the carrier-envelope phase. The CEP is reflected both in the peak height and the position as function of the dimensional raion between amplitude and carrier frequency. For the Gaussian pulse, the impact of the CEP decreases with an increasing number of carrier cycles. As a next step, it would be intriguing to examine the many-particle wave function induced by single-cycle pulses. Noise measurements could be performed in related  experiments as in \cite{Rybka:2016hx,ludwig:2019,Ludwig:2020} to study the effective CEP. As far goal it would be intriguing to access the full many-body wave function, that would allow to extract the energy-time quasiprobablity in analogy to the Wigner function.

\section*{Acknowledgment}

We are grateful for discussions with M. Gedamke, S. Großenbach, H. Kempf, A. Leitenstorfer, and A. Moskalenko. This research was supported by the Deutsche Forschungsgemeinschaft (DFG; German Research Foundation) via SFB 1432 (Project \mbox{No.} 425217212).

\bibliography{Carrier_Envelope_Phase_WB}

\end{document}